# A Catalog of 45,206 Hα Emission-line Stars from LAMOST MRS DR7

Lihuan Yu[1,2], Jiangdan Li[1], Jinliang Wang[3,4], Tongyu He[1,5], and Zhanwen Han[1,2]
[1] International Centre of Supernovae (ICESUN), Yunnan Key Laboratory of Supernova Research, Yunnan Observatories, CAS, Kunming 650216, People's Republic of China; yulihuan@ynao.an.cn, zhanwenhan@ynao.ac.cn
[2] University of Chinese Academy of Sciences, Beijing 100049, People's Republic of China
[3] School of Computational Science and Electronics, Hunan Engineering University, Xiangtan 411104, People's Republic of China
[4] National Astronomical Observatories, Chinese Academy of Sciences, Beijing 100101, People's Republic of China
[5] College of Physics Science and Technology, Hebei University, Baoding 071002, People's Republic of China
Received 2025 August 28; revised 2025 October 28; accepted 2025 October 28; published 2025 December 11

## Abstract

Stars that exhibit prominent emission lines in their spectra are referred to as emission-line stars, encompassing a wide range of stellar types and indicative of intriguing physical properties. The Large Sky Area Multi-Object fiber Spectroscopic Telescope (LAMOST) has released millions of spectra from its Medium-Resolution Survey (MRS). A small fraction of these spectra exhibit emission lines, yet they remain undiscovered and unanalyzed due to being buried in the vast dataset. We have developed a method based on derivative spectroscopy, which provides a novel approach for detecting and identifying emission-line stars by extracting signals from complex backgrounds and estimating spectral line profiles. Applying this method to the Hα spectral line profiles from the LAMOST-MRS Data Release 7, we compiled a catalog of emission-line stars using the second- and third-derivative spectra for automated peak detection. This approach also facilitates the classification of Hα emission-line morphologies through a simplified scheme. The catalog comprises 56,649 spectra with relatively prominent Hα emission lines from 45,206 unique stars, with each emission-line component accompanied by approximate estimates of its wavelength, amplitude, and width. All Hα spectral lines were classified into three morphological classes under a unified classification scheme: single emission peak (83.0%), double emission peaks (5.6%), and P Cygni-type profiles (11.5%), which encompass both P Cygni and inverse P Cygni features. Through cross-referencing with SIMBAD, 39,497 stars represent new emission-line sources discovered in our research.

*Unified Astronomy Thesaurus concepts:* Catalogs (205); Emission line stars (460); Classification systems (253); Stellar spectral lines (1630)

*Materials only available in the online version of record:* machine-readable table

## 1. Introduction

Emission lines arise from the outer regions of stars and can be broadly classified into three main types: (1) Stellar envelopes and outer atmospheres—including expanding envelopes, strong stellar winds, rotating and accretion disks as observed in Be stars, Herbig Ae/Be stars, and T Tauri stars, etc. (A. H. Joy 1942; G. H. Herbig 1960; A. Slettebak 1976; M. Jaschek et al. 1981; U. Finkenzeller & I. Jankovics 1984; J. M. Porter & T. Rivinius 2003; T. Rivinius et al. 2013). (2) Stellar activity—such as flare outbursts, prominences, and starspots as observed in red-dwarf stars, flare stars, red giants, and RS Canes Venatici, etc. (P. R. Wood 1979; S. V. Mallik 1982; C. Cacciari & K. C. Freeman 1983; L. E. Cram et al. 1989; R. D. Robinson et al. 1990; J. C. Hall 2008; Y. Yao et al. 2017). (3) Binary interactions—where processes like mass exchange and accretion flows play a crucial role, as observed in Algol binaries, cataclysmic variables, and symbiotic variables, etc. (R. H. Kaitchuck & R. K. Honeycutt 1982; R. H. Kaitchuck et al. 1985; D. Montes et al. 1996; D. Lin et al. 2012; Y.-J. Zhang et al. 2022). The emission-line characteristics observed in different types of stars reflect their underlying physical processes and evolutionary stages. T. Kogure & K.-C. Leung (2007) classified emission-line stars into four major categories: early-type stars, late-type stars, close binaries, and pre-main-sequence stars. Most of these stars illustrated above exhibit the Hα emission line. Studying spectral line profiles, particularly Hα emission, is essential for understanding the mechanisms of emission-line generation across various stellar types. Furthermore, analyzing spectral line variations provides key observational evidence for investigating the physical conditions and dynamic structures of stellar envelopes and active regions.

In recent years, numerous searches for emission-line stars have been conducted using large-scale spectroscopic surveys. M. Vioque et al. (2020) employed artificial neural networks to compile a catalog of 693 classical Be candidates and 8470 pre-main-sequence candidates, including 1361 Herbig Ae/Be (HAeBe) stars. Their work combined Gaia Data Release 2 (DR2) with data from the Two Micron All Sky Survey, the Wide-field Infrared Survey Explorer, and Hα photometric surveys such as IPHAS and VPHAS+. Additionally, K. Čotar et al. (2020) discovered 10,364 candidate spectra with emission components in the GALAH survey.

Similarly, numerous studies have focused on detecting emission lines in the Large Sky Area Multi-Object fiber Spectroscopic Telescope (LAMOST) spectral data. W. Hou et al. (2016) identified 11,204 spectra of stars with Hα emission based on LAMOST Low-Resolution Survey (LRS) DR2, while P. Škoda et al. (2020) utilized an active learning classification method to identify 1013 spectra of 948 new emission-line candidates from the same dataset. Using LAMOST-LRS DR5 data, R. Anusha et al. (2020) identified

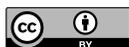

159 classical Ae stars, while B. Shridharan et al. (2021) detected 2716 hot emission-line stars, including 1089 classical Be, 233 classical Ae, and 56 Herbig Ae/Be stars. More recently, Y.-J. Zhang et al. (2022) identified 30,048 spectra with emission lines from LAMOST-LRS DR7, including 201 HAeBe candidates and 5547 CAeBe candidates, while D. Edwin et al. (2024) identified 38,152 late-type emission-line stars from LAMOST-LRS DR6, including 438 infrared excess sources, 4669 post-main-sequence candidates, 9718 Fe/Ge/Ke sources, and 23,264 dMe sources. However, most studies focus on LRS data and have not extended their work to include medium-resolution data, which reveal more detailed spectral line information.

In our previous research (L. Yu et al. 2024), we developed a method based on derivative spectroscopy (DS), which has proven highly effective in identifying weak spectral lines and complex spectral profiles (U. Finkenzeller & R. Mundt 1984). DS was widely utilized by chemists in the 1980s to analyze complex spectral signals, enabling the precise detection and localization of spectral features (T. C. O'Haver et al. 1982). In higher-order derivative spectra, peak widths become narrower, making extreme values—previously obscured by overlapping components—more distinguishable (G. Worthey et al. 1994). Since any spectral profile can be mathematically represented as a function expanded into a polynomial form, the first derivative effectively eliminates the background continuum, thereby enhancing the visibility of spectral features. Compared to other commonly used methods, our DS approach offers broader applicability. For instance, the Gaussian fitting method employed by G. Traven et al. (2015) requires additional Gaussian components as the number of spectral features increases. Machine learning techniques used by M. Vioque et al. (2020), P. Škoda et al. (2020), and Y.-J. Zhang et al. (2022) may also face difficulties in constructing a comprehensive training set, especially for rare or complex spectral line profiles. DS can be universally applied to spectral line profiles of various shapes and component numbers, enabling precise identification of both emission and absorption peaks through multiple derivations. A more detailed comparison of these methods is available in our previous work (L. Yu et al. 2024).

In this study, we catalog the H$\alpha$ emission lines detected by DS in the LAMOST Medium-Resolution Survey (MRS) DR7 dataset and classify them into six common categories: single emission without absorption (1E), single emission with absorption (1EA), double emissions without absorption (2E), double emissions with absorption (2EA), P Cygni profiles (PC), and inverse P Cygni profiles (IPC), providing key parameters such as central wavelength, amplitude, and line width. The structure of this paper is as follows: Section 2 introduces the structure of LAMOST-MRS data and describes the data selection processes, while Section 3 describes the methodology of DS used in this study and the approach to parameter measurement. In Section 4, we present the H$\alpha$ wavelength region adopted in our analysis, along with the method used to eliminate the contamination from nearby spectral lines. In Section 5, we introduce three main classes and six subtypes of H$\alpha$ line profiles along with the criteria used for their classification. In Section 6, we present the catalog of H$\alpha$ emission-line stars, including both previously known sources and newly identified ones based on crossmatching with SIMBAD. Finally, Sections 7 and 8 provide discussion and conclusions, respectively.

## 2. Data Selection

LAMOST is a 4 m Schmidt telescope with a 5° field of view and is equipped with 4000 fibers (G. Zhao et al. 2006, 2012; X.-Q. Cui et al. 2012; A.-L. Luo et al. 2015). The LAMOST-MRS provides spectra at a resolution of $R \sim 7500$, covering the blue (4950–5350 Å) and red (6300–6800 Å) arms, respectively (C. Liu et al. 2020). The LAMOST-MRS test observations started in 2017 September, and the LAMOST-MRS survey began in 2018 October. LAMOST-MRS DR7 includes 1,498,289 FITS files, comprising 11,190,570 single-exposure spectra and 2,906,397 coadded spectra. Each FITS file represents a single observation, containing multiple exposures of spectra at both the red and blue ends during that observation, as well as coadded spectra generated by stacking the multiple exposures. In most cases, the spectral line profiles from multiple exposures taken within a short time frame (a few hours) show no significant variations. This coadded helps to improve the signal-to-noise ratio (SNR) and provides more precise spectral information.

Our focus was solely on the H$\alpha$ line at 6564.6 Å in vacuum, with a bandpass range from 6548 to 6578 Å (G. Worthey et al. 1994; C. Liu et al. 2015), which is located in the red arm of the 1,498,289 coadded spectra. We excluded 341,243 bad spectra due to zero or negative flux values around the H$\alpha$ line, likely caused by missing data in certain wavelength regions or improper instrument calibration. These corrupted spectra were deemed unusable, leaving us with 1,157,046 coadded spectra for further analysis. All spectra were normalized, and cosmic rays were removed using the laspec module (B. Zhang et al. 2020, 2021).

## 3. Derivative Spectroscopy

DS provides an effective approach for separating the emission and absorption components of the H$\alpha$ line in a spectrum. The approach involves analyzing the first, second, and third derivatives of the spectrum to distinguish different spectral components, as illustrated in Figure 1 for both a single emission line (left panels) and a single absorption line (right panels). Each line is constructed using a Gaussian function with amplitude and sigma set to absolute values of 1, with added noise corresponding to an SNR of 20. As the derivative order increases, the corresponding spectra (gray solid lines in the three panels of Figure 1) exhibit enhanced noise, underscoring the necessity of low-pass filtering or smoothing during differentiation. To address this, we applied Gaussian convolution, which leverages convolution properties to obtain various derivatives:

$$\frac{d}{dx}(g \cdot f) = \frac{dg}{dx} \cdot f = g \cdot \frac{df}{dx}. \quad (1)$$

By convolving the spectral line $f$ with the first derivative of a Gaussian function $\frac{dg}{dx}$, one obtains the derivative of the spectrum after Gaussian smoothing $g \cdot \frac{df}{dx}$, i.e., a smoothed derivative spectrum. In this process, a parameter $\sigma_w$ defines the width of the Gaussian convolution kernel, controlling the degree of smoothing applied to the derivative spectra. The solid black lines of Figure 1 with dots depict the derivative spectra smoothed using a Gaussian convolution with a kernel

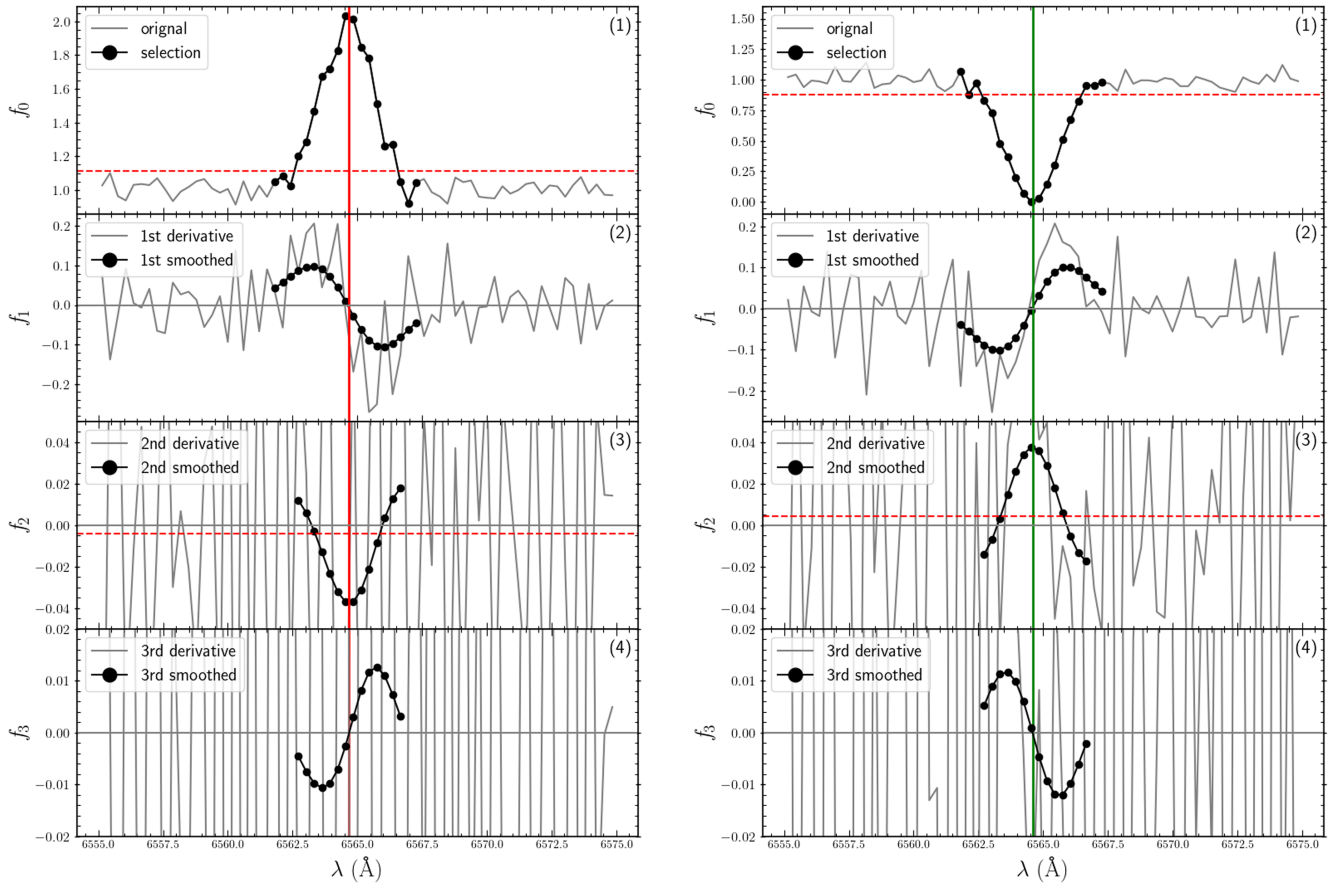

**Figure 1.** The original spectra and their derivatives for a Gaussian function with noise, where the emission line is on the left side, and the absorption line is on the right side. In both figures, panel (1) shows the original spectral line, with red horizontal lines indicating the threshold parameter applied to the original spectrum ($T_0$). Panel (2) presents the first derivative of the spectrum, while panel (3) displays the second derivative, where red horizontal lines mark the threshold parameter for the second derivative ($T_2$). Panel (4) illustrates the third derivative. The gray lines represent the derivative spectra directly calculated, while the black solid lines represent the derivative spectra smoothed using a Gaussian kernel with a width ($\sigma_w$) of three. The red vertical solid line indicates the central wavelength of the emission line, and the green vertical solid line marks absorption lines.

width $\sigma_w$ of 3, and the kernel is defined by

$$G(x) = \frac{1}{\sqrt{2\pi}\,\sigma_w} e^{-\frac{x^2}{2\sigma_w^2}}. \quad (2)$$

In Figure 1, an emission peak (on the left) in the original spectrum corresponds to a zero crossing in the declining part of the first derivative, a negative peak in the second derivative, and a zero crossing in the rising part of the third derivative. We applied thresholds to both the original spectrum and its second-order derivative—denoted as $T_0$ and $T_2$, respectively—to identify spectral peaks (values exceeding these thresholds were considered signals rather than noise). These thresholds are illustrated by the red horizontal dashed lines in Figure 1. The filtered regions in the second-order derivative (black solid lines with points exceeding $T_2$ in Figure 1) help determine whether an emission peak is present, while filtering in the original spectrum (black solid lines with points exceeding $T_0$) indicates whether the peak lies above or below the continuum. In addition, zero crossings in the first- and third-order derivatives (indicated by red vertical solid lines in Figure 1) are used to precisely locate the emission line. As shown in the right panel of Figure 1, the detection of absorption lines was carried out in a manner similar to that used for emission lines.

The thresholds ($T_0$ and $T_2$) are determined based on the mean and standard deviation of the surrounding continuum. To ensure accurate estimation, the surrounding continuum is taken as a 5 Å region on both sides of the H$\alpha$ line. This range is chosen because noise-induced dispersion is more strongly correlated with the spectral lines in closer regions, leading to a more reliable determination of the threshold.

As shown in Figure 1, the parameters $T_0$ and $T_2$ differ slightly between the detection of emission and absorption lines. We take emission lines as an example to illustrate how these two parameters are determined. For emission lines, $T_0$ is defined as the mean flux of the surrounding continuum plus 3 times its standard deviation:

$$T_0 = f_0 + 3\sigma_0, \quad (3)$$

where $f_0$ and $\sigma_0$ denote the mean flux and standard deviation of the original spectrum, respectively. This threshold effectively filters out most of the noise, as 99.7% of values in a Gaussian distribution are expected to fall within three standard deviations of the mean. In this study, an emission peak is considered to lie above the continuum if its extremum is higher than $f_0 + 3\sigma_0$.

The parameter $T_2$, determined in a manner similar to the threshold for the original spectrum. The parameter $T_2$ determines whether a peak is present in our detection results and whether it exhibits emission or absorption. For emission lines, it is defined as the mean of the surrounding continuum in the second derivative, minus a fixed multiple of its standard

deviation:

$$T_2 = f_2 - k\sigma_2, \qquad (4)$$

where $f_2$ and $\sigma_2$ denote the mean and standard deviation of the second derivative, respectively, and $k$ is a fixed coefficient, which is chosen by following calculation. We generated synthetic H$\alpha$ emission profiles using a Gaussian function with added white noise. The amplitude range of the Gaussian function is related to the SNR of the synthetic spectra. Specifically, the amplitude is defined as a multiple of the inverse SNR (i.e., the noise level sigma), with the multiplier ranging from 3 to $10^2$. This wide range is chosen to cover a broader variety of line profile shapes, allowing for a more accurate determination of the optimal value of $k$. Additionally, the width of the spectral lines, modeled using a Gaussian profile, is randomly selected between 0 and 5 Å. This range effectively covers the majority of H$\alpha$ line widths observed in the LAMOST-MRS DR7 dataset. The SNRs of the synthetic spectra are sampled from 0 to 100. Except for a few spectra with very high SNR, this range basically covers the entire SNR distribution of the LAMOST-MRS dataset.

Each test sample consists of 10,000 synthetic H$\alpha$ emission lines and 10,000 noise-only spectra at the same resolution. This setup is used to evaluate the precision and recall of the spectral profile classification under different values of $k$. Precision is the proportion of true emission lines among all lines detected by our method, while recall is the proportion of emission lines detected by our method among all true emission lines. The precision and recall for $k$ values ranging from 1 to 5 are shown in Figure 2. As $k$ increased, precision improved while recall decreased. We ultimately set $T_2$ to the mean of the surrounding continuum plus or minus 4.5 times its standard deviation, at which point the precision reached approximately 95%, and the recall was approximately 98%. A peak is classified as an emission component if its second derivative lies below $f_2 - 4.5\sigma_2$, as shown in Figure 1.

For absorption lines, the determination of the two parameters is similar to the procedure described above, with $T_0$ adjusted to $f_0 - 3\sigma_0$ and $T_2$ adjusted to $f_2 + 4.5\sigma_2$. If the second-order derivative exceeds $T_2$, the presence of an absorption component is assumed, and if this absorption lies below $T_0$, the component is considered to be below the continuum.

$\sigma_w$ must be carefully chosen to balance the trade-off between smoothing and peak detection accuracy. If $\sigma_w$ is too large, excessive smoothing may obscure closely spaced peaks, while if it is too small, numerical noise from successive derivatives could interfere with detection. A larger convolution kernel is more effective for identifying broader peaks, whereas a smaller kernel is better suited for detecting narrow features. To optimize detection across a large number of spectra, multiple values [2, 3, 4, 5, 6] of $\sigma_w$, which were uniformly selected within the range of 2–6 Å. The line profile was estimated using a cross-validation approach, and a peak was considered to be reliably detected only if it was consistently identified under more than two different smoothing widths. This strategy ensures that the most suitable smoothing scale for spectral line detection can be adaptively determined for diverse spectral conditions.

### 3.1. Parameters of a Line's Peaks

Through the DS-based detection process, the number of peaks ($N_p$) within a spectral line profile can be readily determined, along with their basic characteristics. Peaks where the second derivative falls below $f_2 - 4.5\sigma_2$ are identified as emission features, while those above $f_2 + 4.5\sigma_2$ are classified as absorption features. Based on the characteristics of the derivative spectra, we can preliminarily estimate the basic information of each peak, including amplitude $a$, central wavelength $\lambda$, and width $w$. As shown in Figure 1, the red vertical line on the left panels and the green vertical line on the right panels indicate the central wavelengths of the emission and absorption lines, respectively. These wavelengths correspond precisely to the zero-crossings of the first derivative, the extrema of the second derivative, and the zero-crossings of the third derivative. Since higher-order derivatives are capable of separating complex components of spectral lines, some overlapping components that are distinguishable in the third-order derivative are not clearly separated in the first-order derivative. Moreover, the narrowing of spectral line features in higher-order derivatives enables more precise localization of peak extrema, thereby improving the accuracy of central wavelength detection. Therefore, the value of $\lambda$ is determined based on the third-order derivative, while the first-order derivative serves as a verification tool.

At the same time, there is one positive and one negative peak on either side of the zero-crossings in both the first and third derivatives. The distance (in wavelength) between these two peaks is closely related to the line width of the original spectrum. Assuming that each peak in the H$\alpha$ line follows a Gaussian distribution such as Figure 1. Analytical calculations show that the distance between the two peaks in the first derivative corresponds exactly to the standard deviation ($\sigma$) of the Gaussian function, while the distance between the two peaks in the third derivative is approximately $0.74\sigma$. For the same reason as the choice of the central wavelength ($\lambda$), we adopt the distance between the peaks flanking the zero-crossing in the third-order derivative as the width ($w$) of each identified component.

Finally, the amplitude ($a$) of each component is obtained by substituting the central wavelength ($\lambda$) into the original spectrum, which is achieved through linear interpolation. Note that the width parameter is more accurate for peaks that follow a Gaussian distribution, while the amplitude $a$ only represents the flux at the central wavelength.

Since the smoothing interpolation process does not provide measurement-like uncertainties, we adopted a statistical approach to estimate the parameter errors. As an example, we consider a spectrum with an SNR of 20. White noise with a standard deviation of 1/20 is randomly added 1000 times. This process generates 1000 realizations of the same spectral line under this SNR condition. The uncertainty of each parameter for this spectral line is taken as the standard deviation of the corresponding parameter values obtained from the 1000 simulated detections. For example, the uncertainty of the wavelength, $\sigma_\lambda$, is given by

$$\lambda_{\text{err}} = \sqrt{\frac{1}{N-1}\sum_{i=1}^{N}(\lambda_i - \bar{\lambda})^2}, \qquad (5)$$

where $N = 1000$, $\lambda_i$ represents the wavelength measured in the $i$th iteration, and $\bar{\lambda}$ is their mean value. The uncertainties of all measured parameters are obtained by applying the same statistical procedure described above. Due to the large number of spectra, it is difficult to estimate the uncertainties of each parameter for every spectrum using this approach. After analyzing a subset of spectra, we found that the estimated uncertainties are highly correlated with SNR. We thus employed the following procedure to estimate uncertainties for the full sample.

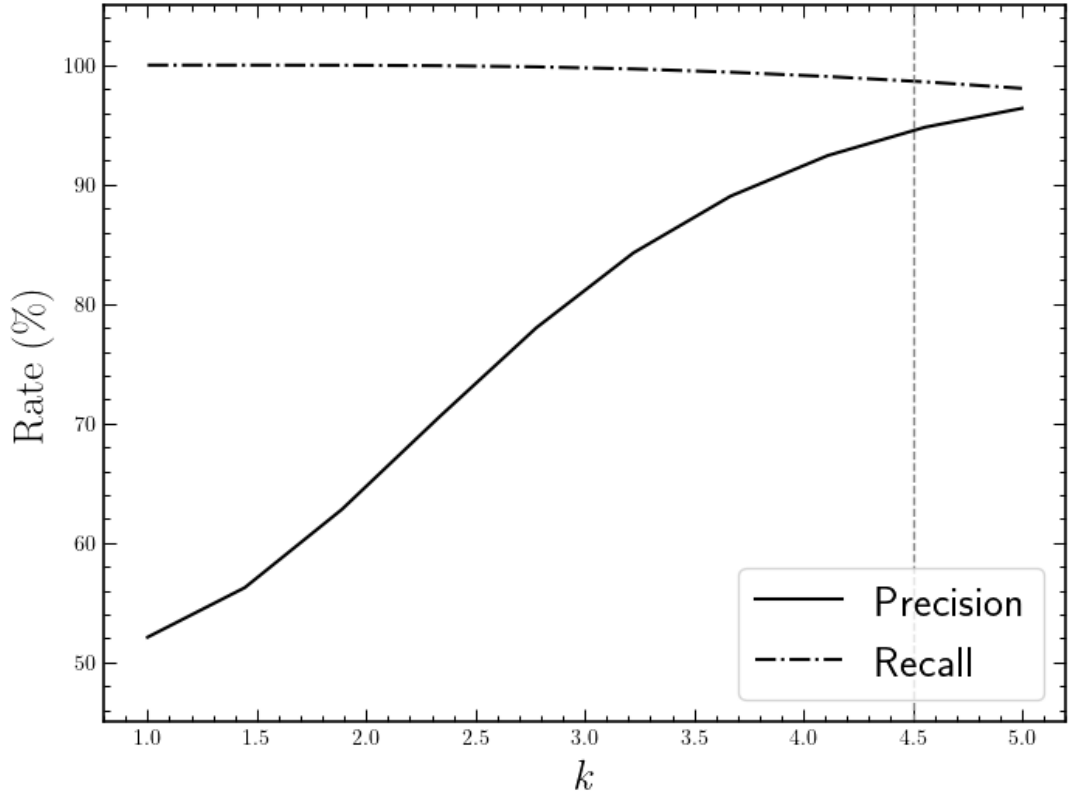

**Figure 2.** Detection rate with different parameter settings for $k$. The selected value $k = 4.5$ is marked with a gray vertical line. The solid line represents precision, defined as the ratio of correctly detected emission lines to all detected lines. The dashed line represents recall, defined as the ratio of detected emission lines to the total number of emission lines in the sample.

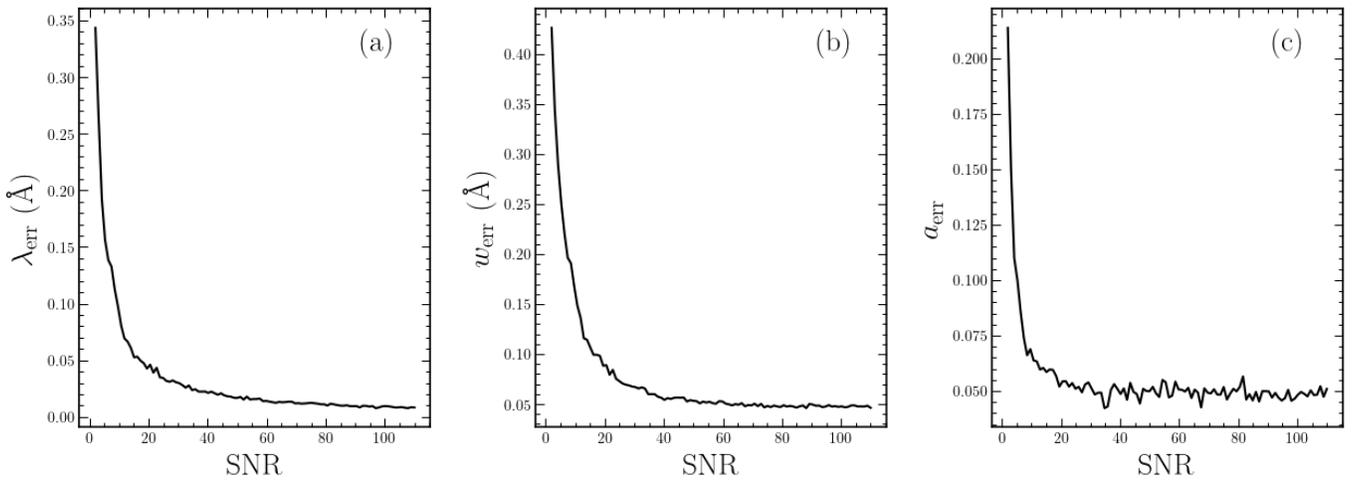

**Figure 3.** The relationship between $\lambda_{err}$, $w_{err}$, $a_{err}$, and SNR. Panel (a) represents the change in the error of the central wavelength with respect to SNR, while panels (b) and (c) represent the errors in width and amplitude, respectively.

For the artificially synthesized H$\alpha$ lines mentioned earlier, we randomly sampled 1000 spectra at each SNR, calculating the standard deviation of the residuals between the true parameter values ($\lambda$, $a$, $w$) of the samples and the derivative spectrum method measurements. This standard deviation is then used as an estimate of the parameter's error at the corresponding SNR. Figure 3 shows the relationship between the error and SNR. All parameters presented in this paper are accompanied by error estimates derived from the relationship between SNR and error. It is important to note that these are not the errors calculated directly from the measured spectra, but rather estimated values of the error.

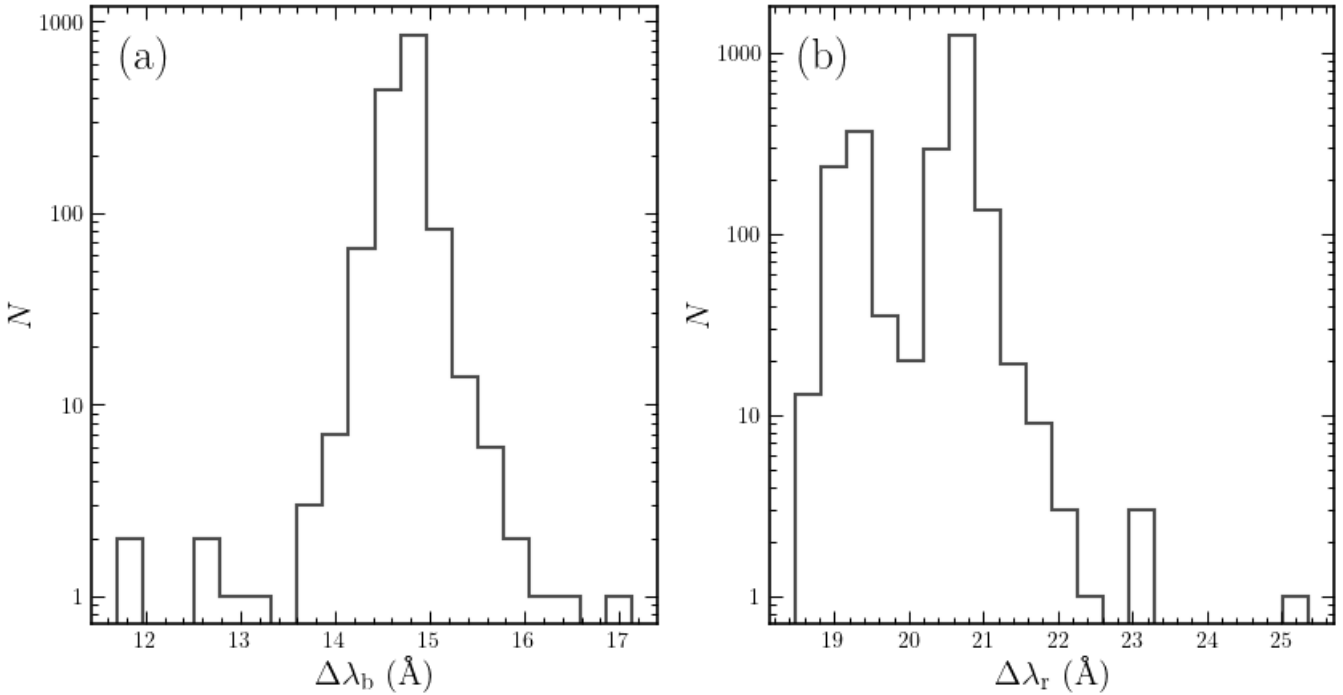

**Figure 4.** Distributions of wavelength separations between Hα and [N II] of spectra of H II regions. Panel (a) represents the distribution of $\Delta\lambda_b$, and panel (b) represents $\Delta\lambda_r$, respectively.

## 4. Nebulae

Nebulae—interstellar clouds of dust, gas, and plasma—significantly influence the observed spectra of nearby sources. When a star within a nebula emits enough ultraviolet photons, it ionizes the surrounding medium, producing characteristic nebular emission lines such as Hα, [N II], [S II], and [O III]. If a source is projected within a nebular region, its spectrum may be contaminated by these lines, making it difficult to determine whether the emission is intrinsic to the source or of nebular origin. To mitigate this issue, we employed the following approach to distinguish the origin of this portion of the Hα emission line.

Based on the crossmatch between our sample and the H II regions cataloged by L. D. Anderson et al. (2020), 0.6% of the spectra of emission-line stars are projected onto H II regions and are flagged as H II contamination in our catalog. In addition, we used DS to examine the nearby [N II] lines at vacuum wavelengths of 6549.8 and 6585.6 Å. The selected wavelength bands for [N II] are 6547–6553 Å and 6582–6588 Å. If emission lines are also present in the [N II] wavelength region, it suggests that the Hα emission line may originate from nebular activity. Approximately 47.7% of the Hα emission-line spectra in our sample show [N II] emission lines, which are flagged as [N II] contamination.

Under the assumption that both Hα and [N II] emission originate from the same nebula and share the same line-of-sight velocity, the wavelength separations between Hα and the [N II] lines at $\lambda_b = 6549.8$ Å and $\lambda_r = 6585.6$ Å are fixed, approximately $\Delta\lambda_b = 14.8$ Å and $\Delta\lambda_r = 21.0$ Å, consistent with their laboratory values. However, statistics for single-emission-line spectra within H II regions show that the mean of $\Delta\lambda_b$ is 14.72 Å with a standard deviation of 0.27 Å, and the mean of $\Delta\lambda_r$ is 20.21 Å with a standard deviation of 0.74 Å. The distributions are shown in Figure 4: panel (a) represents $\Delta\lambda_b$ and panel (b) represents $\Delta\lambda_r$. We selected the range of $\Delta\lambda_b$ as 13.91–15.53 Å. The distribution in panel (b) shows two distinct concentrations. At around 19.2 Å, the separation between the Hα and N II emission lines deviates significantly from the fixed value of 21 Å, suggesting a line-of-sight velocity difference between their emission regions. This indicates that the Hα emission may originate from the star itself. The range of $\Delta\lambda_r$ is selected based on the peak at 20.6 Å as 20–21.38 Å. Among the sources flagged as [N II] contamination, 30% have both $\Delta\lambda_r$ and $\Delta\lambda_b$ within the selected ranges, suggesting a nebular origin of the Hα emission. Meanwhile, 50% satisfy the condition for only one of the two sides, and 20% satisfy neither condition.

In our catalog, the columns H II and [N II] are used to mark these sources. A "Y" in H II indicates that the emission-line star is projected onto an H II region, while a "Y" in [N II] denotes the presence of [N II] emission lines. The latter suggests that the Hα emission may originate from nebular activity. An "N" indicates the opposite case, implying that the Hα emission does not originate from a nebula and may instead be stellar in origin. For each spectrum exhibiting [N II] emission lines, we provide the wavelengths of both the red and blue components. This information enables users to preliminarily assess whether the emission arises from nebular activity or from sky-line contamination. Due to the complexity of a small fraction of [N II] line profiles, less than 1% of the wavelengths may be inaccurately determined.

## 5. Morphological Classification

The spectra were categorized into three major types: single emission profiles, double emission profiles, and P Cygni-type profiles. Each of these three primary types is further divided into two distinct subtypes, resulting in a total of six morphological categories. Representative Hα spectra for each of these six categories, drawn from the LAMOST-MRS dataset, are illustrated in Figure 5. All categories are consistently labeled using a unified symbolic notation, in which the letters E and A denote the emission and absorption components, respectively. The classification scheme for all six categories is listed in Table 1, followed by detailed explanations and representative examples for each category.

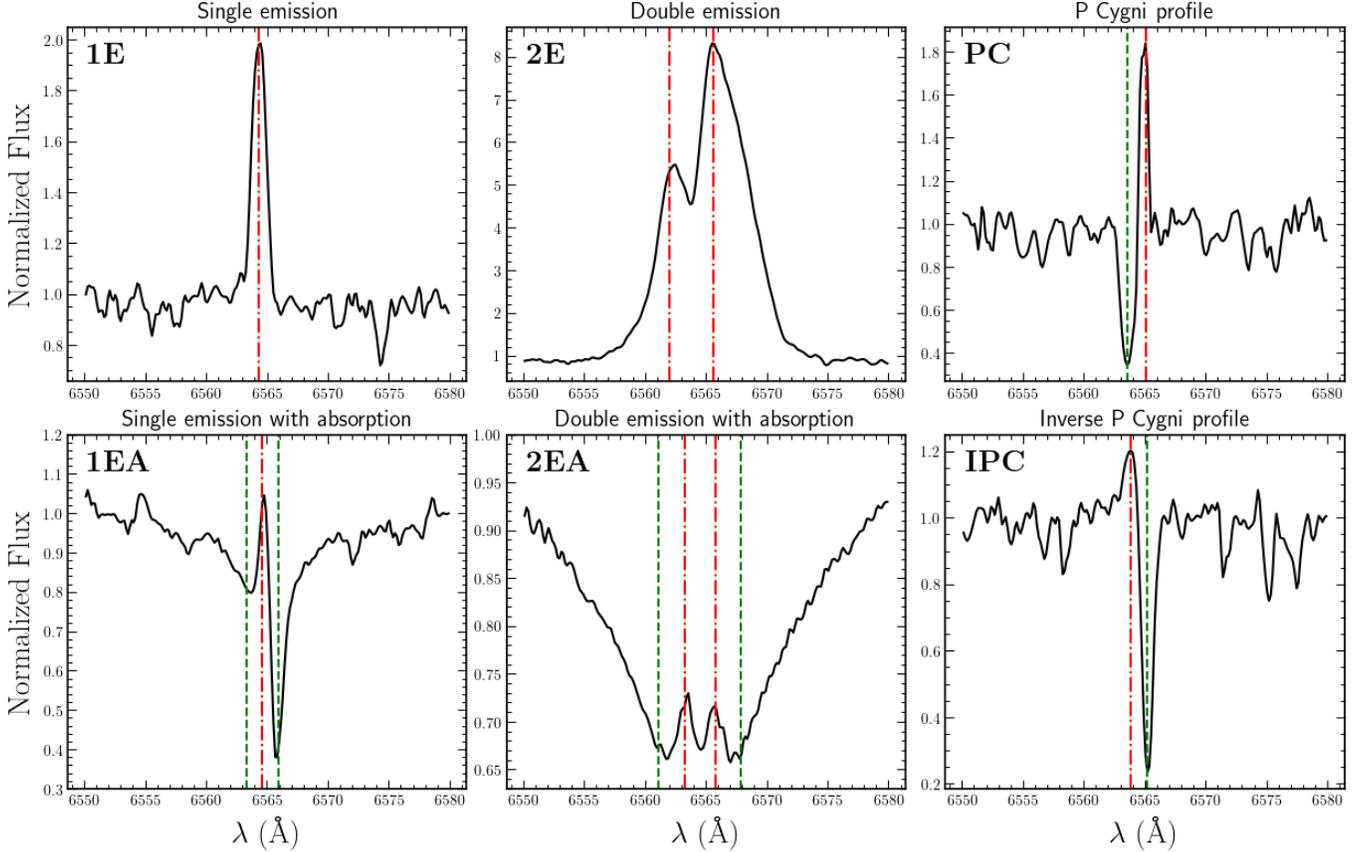

**Figure 5.** Examples of six categories (1E: J110243.48+554448.2, 1EA: J063453.62+092617.3, 2E: J041912.84+282932.8, 2EA: J064208.72+225205.7, PC: J020131.75+384248.6, and IPC: J082231.49+202957.2) of three major types with H$\alpha$ emission lines. The red vertical dashed–dotted lines indicate the central wavelength of emission lines, while the green vertical dashed lines indicate the central wavelength of absorption lines.

**Table 1**
Categories with Corresponding Selection Criteria

| Class | Subtype | Flags | Criteria 1 | Criteria 2 | Criteria 3 |
|---|---|---|---|---|---|
| Single emission profile | without absorption | 1E | $N_p = 1$ | $f_2(\lambda_1) < T_2$ | $a_1 > T_0$ |
|  | with absorption | 1EA | $N_p = 3$ | $f_2(\lambda_2) < T_2, f_2(\lambda_{1,3}) > T_2$ | $a_{1,3} < T_0$ |
| Double emission profile | without absorption | 2E | $N_p = 2$ | $f_2(\lambda_{1,2}) < T_2$ | $a_{1,2} > T_0$ |
|  | with absorption | 2EA | $N_p = 4$ | $f_2(\lambda_{1,4}) > T_2, f_2(\lambda_{2,3}) < T_2$ | $a_{1,4} < T_0$ |
| P Cygni-type profile | P Cygni profile | PC | $N_p = 2$ | $f_2(\lambda_1) > T_2, f_2(\lambda_2) < T_2$ | $a_1 < T_0, a_2 > T_0$ |
|  | Inverse P Cygni profile | IPC | $N_p = 2$ | $f_2(\lambda_1) < T_2, f_2(\lambda_2) > T_2$ | $a_1 > T_0, a_2 < T_0$ |

1. *Single emission profiles include two categories:* without absorption (1E) and with absorption (1EA). Both of them showed one and only one emission line.

   The 1E profile exhibits only a single emission component. For 1E, the second derivative must be below $T_2$, and the amplitude $a$ of the peak must be higher than $T_0$.

   The 1EA profile includes a broad, shallow absorption component superimposed on the emission line, resulting in an absorption peak on each side of the emission line. In the derivative spectrum, we observe three peaks: two absorption peaks and one emission peak. The second derivative of the central emission peak must be below $T_2$, while the second derivatives of the absorption peaks must exceed $T_2$. Additionally, the amplitudes of the absorption peaks, $a_1$ and $a_3$, must be below $T_0$.

2. *Double emission profiles, like single emission profiles, also include two categories:* without absorption (2E) and with absorption (2EA). Both 2E and 2EA show two distinct emission peaks. Note that in these two categories, an absorption peak is expected to appear between the double emission lines. However, for some spectra, the double emission lines overlap so strongly that the central absorption feature is not visible. Since the parameters of this peak do not affect our classification of the line profiles, it is not considered in this work.

   The 2E profile contains only the two emission components. The second derivative of two emission peaks must be below $T_2$, and the amplitudes ($a_1$ and $a_2$) must exceed $T_0$.

Table 2
The Number of Stars in Each Subtype

| Subtypes | Flags | All Emission-line Stars | | | Without [N II] Line | | |
|---|---|---|---|---|---|---|---|
| | | No. Obj. | No. Spec. | Fraction (%) | No. Obj. | No. Spec. | Fraction (%) |
| Single emission profiles | 1E | 13,069 | 15,370 | 27.1 | 6167 | 7379 | 24.1 |
| | 1EA | 20,274 | 31,627 | 55.8 | 16,289 | 19,120 | 62.5 |
| Double emission profiles | 2E | 2214 | 3015 | 5.3 | 1434 | 2032 | 6.6 |
| | 2EA | 90 | 102 | 0.2 | 90 | 102 | 0.3 |
| P Cygni-type profiles | PC | 3855 | 4180 | 7.4 | 1040 | 1099 | 3.6 |
| | IPC | 1950 | 2355 | 4.2 | 784 | 867 | 2.8 |
| All profiles | ... | 45,206 | 56,649 | 100 | 25,490 | 30,599 | 100 |

The 2EA profile features a broad, shallow absorption line similar to that in 1EA, with the two emission peaks embedded within the absorption profile. Four peaks are observed at the derivative: the second derivatives two emission peaks must be below $T_2$, while the second derivatives of the two absorption peaks must exceed $T_2$. The amplitudes of the absorption peaks ($a_1$ and $a_4$) must be below $T_0$.

3. *P Cygni-type profiles are further divided into two categories:* P Cygni profiles (PCs) and Inverse P Cygni profiles (IPCs). Both PC and IPC show an emission line and a shifted absorption line.

PCs feature an emission line with a blueshifted absorption component. The second derivative of the emission component lies below $T_2$, while that of the absorption component lies above it. The emission peaks $a_1$ are required to exceed $T_0$, whereas the absorption peak $a_2$ must fall below $T_0$.

IPCs exhibit an emission line accompanied by a redshifted absorption feature. The second derivative of the emission component lies below $T_2$, while that of the absorption component lies above it. The absorption peaks $a_1$ are required to lie below $T_0$, whereas the emission peaks $a_2$ must exceed $T_0$.

## 6. Results

### 6.1. Catalog

Among the 1,037,912 spectra analyzed, only 120,708 spectra from 86,364 objects exhibit potential emission components of varying strengths. To ensure the reliability of the emission-line star catalog, we applied the selection criteria listed in Table 1, ultimately retaining 45,206 objects with strong and distinct emission lines that fall into six subtypes. The number and fraction of stars in each subtype are summarized in Table 2. In the "All Emission Line Stars" part of Table 2, "No. Obj." refers to the total number of stars with Hα emission lines, and "No. Spec." denotes the total number of corresponding spectra. For the "Without [N II] Line" subset, "No. Obj." and "No. Spec." show the same values but only for spectra exhibiting Hα emission lines without accompanying [N II] lines.

In our sample, single emission profiles constitute the largest group, representing approximately 83.0% of the total. Specifically, the 1E type makes up 27.1%, while 1EA accounts for 55.8%. This may be attributed to the broad range of origins and diverse formation mechanisms associated with single-peaked emission lines. Double-peaked emission profiles are the rarest in our sample, accounting for 5.6% in total, 5.3% for the 2E subtype, and only 0.2% for 2EA. This rarity is likely due to the stringent conditions required for their formation: the presence of a circumstellar disk, and the star being mainly observed edge-on, or the system having a high inclination (T. Rivinius et al. 2013). P Cygni-type profiles account for approximately 11.5% of the sample (7.4% for PC and 4.2% for IPC), but over 70% of these spectra are affected by nebular contamination. When a star lies within a nebular region, and there is a radial velocity difference between the two, it is easy for profiles resembling PC and IPC types to emerge.

All 45,206 objects, corresponding to 56,649 spectra with detected Hα emission lines, have been included in our catalog (Table 3), which is available in machine-readable format and in the main astronomical catalog repository, CDS VizieR. The catalog provides a detailed description of each object and its spectral features. The column "Designation" lists the LAMOST identifier for each source, while "Simbad ID" gives the matched identifier from the Simbad database, where available. Each spectrum is recorded by its LAMOST-MRS FITS file name under the column "Specname." The morphological classification of the Hα emission-line profile is recorded in the column "Flags(Hα)"; the stellar type, if available from SIMBAD, is listed in the "Type" column. Detailed descriptions of these types are provided in the next section. To indicate the potential origin of the Hα emission, we introduce two flags, H II and [N II], which are assigned based on whether the star is projected onto an HII region and whether [N II] emission lines are present in the spectrum, respectively. A value of "Y" indicates that the star is located within an H II region, or that [N II] lines are detected. This suggests a likely nebular origin for the Hα emission. In contrast, a value of "N" suggests that the emission is likely of stellar origin. For spectra exhibiting [N II] emission lines, $\lambda_{\mathrm{[NII]B}}$ denotes the central wavelength of the blue-side [N II] line, and $\lambda_{\mathrm{[NII]R}}$ denotes that of the red-side line, which facilitates users in further determining whether the emission line originates from a nebula.

The parameters of individual components of profiles are also provided. The columns $\lambda_1$, $\lambda_2$, $\lambda_3$, $\lambda_4$ list the central wavelengths of up to four emission peaks in each Hα profile, ordered from the short to the long wavelength. The uncertainty in the central wavelength, denoted as $\lambda_{\mathrm{err}}$, is estimated based on synthetic spectra with similar SNR. Similarly, the peak

Table 3
Catalog of Hα Emission-line Stars

| Num | Quantity | Column | Format | Units | Description |
|---|---|---|---|---|---|
| 1 | Designation | desig | Char(19) | ⋯ | LAMOST designation |
| 2 | Simbad ID | simbad_id | Char | ⋯ | Source ID from SIMBAD |
| 3 | R.A. | radeg | Float | deg | R.A. at epoch 2000.0 (ICRS) |
| 4 | DEC. | decdeg | Float | deg | Decl. at epoch 2000.0 (ICRS) |
| 5 | Specname | specname | Char | ⋯ | FITS file name of LAMOST-MRS |
| 6 | Flags(Hα) | flags | Char | ⋯ | Morphological categories of Hα lines |
| 7 | Type | Obj_type_literature | Char | ⋯ | Object classification reported in the literature |
| 8 | H II | H II | Char(1) | ⋯ | Projected on H II regions: Y = Yes, N = No |
| 9 | [N II] | [N II] | Char(1) | ⋯ | Presence of [N II] emission: Y = Yes, N = No |
| 10 | $\lambda_{[NII]B}$ | lam_[NII]_B | Float | Å | Central wavelength of the [N II] line at blue side |
| 11 | $\lambda_{[NII]R}$ | lam_[NII]_R | Float | Å | Central wavelength of the [N II] line at red side |
| 12 | $\lambda_1$ | lam_1 | Float | Å | Central wavelength of the first peak |
| 13 | $\lambda_2$ | lam_2 | Float | Å | Central wavelength of the second peak |
| 14 | $\lambda_3$ | lam_3 | Float | Å | Central wavelength of the third peak |
| 15 | $\lambda_4$ | lam_4 | Float | Å | Central wavelength of the fourth peak |
| 16 | $\lambda_{err}$ | lam_err | Float | Å | Statistical error of the central wavelength |
| 17 | $a_1$ | a_1 | Float | ⋯ | Amplitude of the first peak |
| 18 | $a_2$ | a_2 | Float | ⋯ | Amplitude of the second peak |
| 19 | $a_3$ | a_3 | Float | ⋯ | Amplitude of the third peak |
| 20 | $a_4$ | a_4 | Float | ⋯ | Amplitude of the fourth peak |
| 21 | $a_{err}$ | a_err | Float | ⋯ | Statistical error of amplitude |
| 22 | $w_1$ | w_1 | Float | Å | Width of the first peak |
| 23 | $w_2$ | w_2 | Float | Å | Width of the second peak |
| 24 | $w_3$ | w_3 | Float | Å | Width of the third peak |
| 25 | $w_4$ | w_4 | Float | Å | Width of the fourth peak |
| 26 | $w_{err}$ | w_err | Float | Å | Statistical error of width |
| 27 | SNR | SNR | Float | ⋯ | Signal-to-noise ratio |

(This table is available in its entirety in machine-readable form in the online article.)

amplitudes relative to the continuum are given in the columns $a_1$, $a_2$, $a_3$, $a_4$, with a corresponding error $a_{err}$ derived through the same statistical approach. The widths of the emission components are provided in $w_1$, $w_2$, $w_3$, $w_4$, and the column $w_{err}$ gives the associated width uncertainties. However, for the 1EA category, approximately 44.6% of the Hα emission lines are relatively narrow and weak, and are superimposed on broad and deep absorption features. This makes their widths difficult to estimate, and, consequently, the $w_2$ parameter is not provided in the catalog. The SNR for each spectrum is also provided in the column SNR.

### 6.2. Newly Discovered Stars with Hα Emission Lines

There are 39,497 newly discovered Hα emission-line stars in our sample that have not been previously listed as emission-line stars. Among them, 29,373 objects lack any counterpart in the SIMBAD[6] database (matched within 3″), and 10,124 are listed as "star" in SIMBAD but had not been previously recognized as emission-line stars. Among them, type 1E accounts for 14.3%, type 1EA for 76.1%, type 2E for 3.2%, type 2EA for 0.2%, type PC for 3.2%, and type IPC for 3.0%. The vast majority exhibit relatively clear and well-defined profile structures. The examples presented in Figure 5 are all newly detected Hα emission-line stars, which are currently classified as "Star" in the SIMBAD database.

Through crossmatching with the SIMBAD database, we identified a total of 15,833 sources that have been previously observed or studied. After excluding the 10,124 sources identified as normal stars and the 794 recognized as emission-line stars, 4915 additional objects remain with more detailed stellar classifications. Specific classifications include 1522 binaries, 977 variables, 720 evolved stars, 463 pre-main-sequence stars, and 256 main-sequence stars (P. R. Woźniak et al. 2004; L. Venuti et al. 2015; A. N. Heinze et al. 2018; M. Kounkel et al. 2019; M. Shetrone et al. 2019; X. Chen et al. 2020; L.-Y. Zhang et al. 2020; L. Wang et al. 2022). Additionally, 977 other sources were identified based on unusual chemical signatures, extreme mass values (either exceptionally large or small), distinctive spectral properties (such as UV emission, infrared, and X-ray, etc.), or associations with the interstellar medium or star clusters. The detailed categories of stars included within each major group are described below:

1. *Binaries.* The most common type in our sample is eclipsing binaries, comprising 754 confirmed sources and 11 candidates. Subsequently, 497 sources are identified as double-lined spectroscopic binaries, 129 as RS Canum Venaticorum stars, 4 as cataclysmic variables with 3 candidates, 2 as ellipsoidal variables, 1 as an X-ray binary, and 121 as unclassified binary or multiple star systems (D. I. Hoffman et al. 2009; J. Alfonso-Garzón et al. 2012; D. Lin et al. 2012; P. Klagyivik et al. 2013; D. M. Bramich et al. 2014; A. J. Drake et al. 2014; H. A. Kobulnicky et al. 2014; D. J. Armstrong et al. 2015; X. Chen et al. 2020).

2. *Variables.* The sample comprises 146 BY Draconis variables, 94 rotating variables, 98 eruptive variables, 98

---

[6] https://simbad.u-strasbg.fr/simbad/sim-fbasic

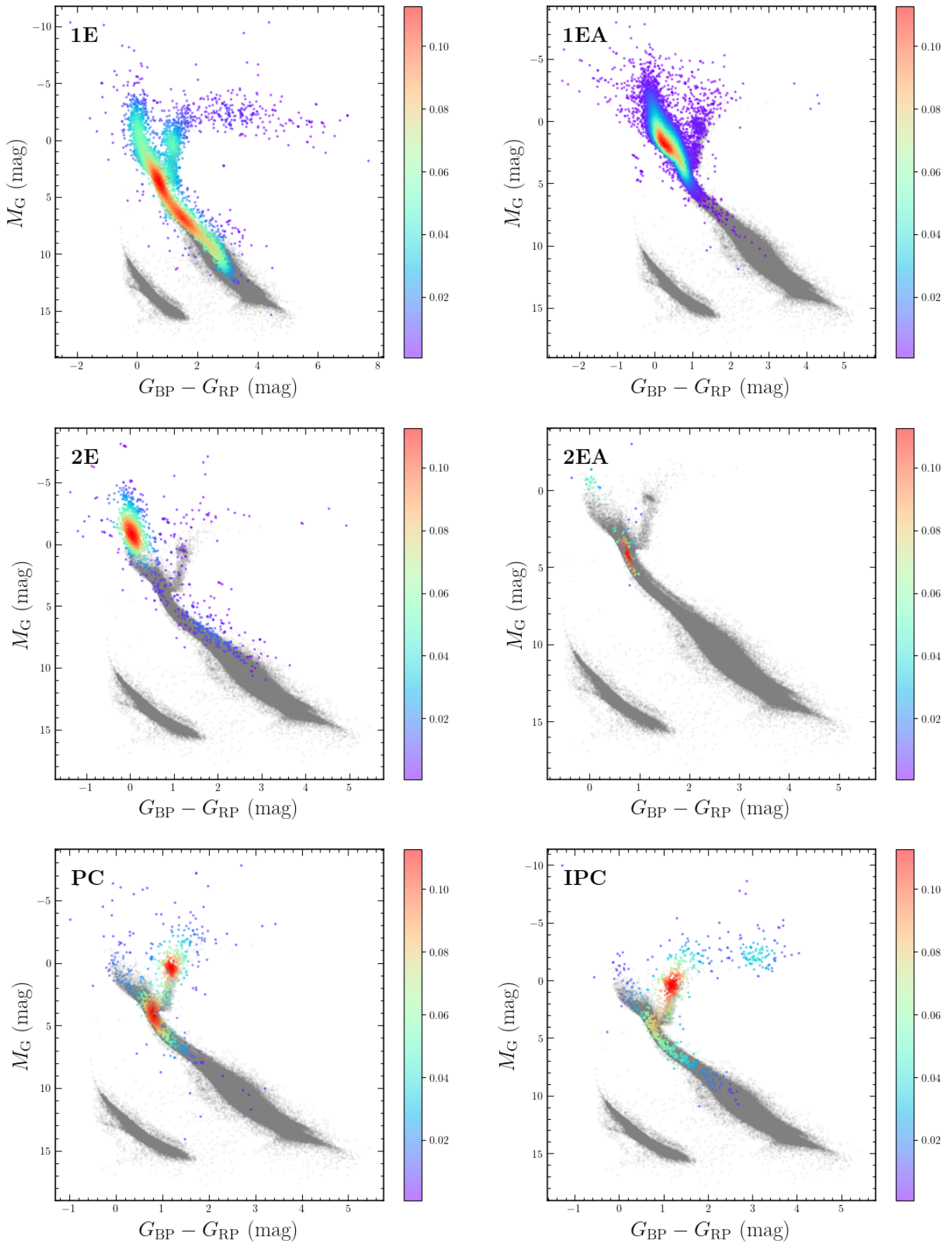

**Figure 6.** H-R diagrams for stars with H$\alpha$ emission lines across six different labels (1E, 1EA, 2E, 2EA, PC, and IPC). Gray dots indicate background stars within 100 pc from the Gaia catalog. Colored points show the density distribution of emission-line stars selected from our sample.

  Orion variables, 101 pulsating variables, 86 Delta Scuti variables, and 354 general variables (E. W. Gottlieb & W. Liller 1978; L. A. Balona & W. A. Dziembowski 2011; T. Reinhold et al. 2013; A. McQuillan et al. 2014; D. J. Armstrong et al. 2015; N. N. Samus' et al. 2017; A. N. Heinze et al. 2018; G. Michalska 2019; L.-Y. Zhang et al. 2020).

3. *Evolved stars.* This category features the most diverse stellar types, including 214 long-period variables and 61 candidates, 152 red giant branch stars and 1 candidate, 76 Mira variables, 49 horizontal branch stars, 48 carbon stars, 23 RR Lyrae variables, 16 Cepheids, 11 hot subdwarfs and 9 candidates, 8 S-type stars and 36 candidates, 7 asymptotic giant branch (AGB) stars and 1 candidate, 5 Type II Cepheids, 2 RV Tauri variables, and 1 post-AGB star (E. J. Totten & M. J. Irwin 1998; C. Akerlof et al. 2000; P. R. Woźniak et al. 2004; M. Ness et al. 2016; Z. Lei et al. 2018; T. Jayasinghe et al. 2019; M. Shetrone et al. 2019; P. Gaulme et al. 2020; J. Chen et al. 2023).

4. *Pre-main-sequence stars.* The sample comprises 187 young stellar objects and 117 candidates, 152 T Tauri stars and 2 candidates, and 5 Herbig Ae/Be stars (L. Cieza et al. 2007; S. T. Megeath et al. 2012; L. Venuti et al. 2015; J. N. Cottle et al. 2018; M. Kounkel et al. 2019).

5. *Main-sequence stars.* The sample comprises 134 Be stars and 81 candidates, 26 gamma Dor variables, and 10 blue stragglers and 5 candidates (S. D. Chojnowski et al. 2015; A. M. Geller et al. 2015; Z. Liu et al. 2022; L. Wang et al. 2022).

## 7. Discussion

By crossmatching with Gaia, we obtained 23,021 objects exhibiting intrinsic Hα emission lines without [N II] emission lines for further analysis. To correct for interstellar extinction, we applied extinction coefficients from Gaia (L. Casagrande & D. A. VandenBerg 2018) and Bayestar19 (G. M. Green et al. 2019), using the extinction relations of J. A. Cardelli et al. (1989) and E. L. Fitzpatrick (1999). Compared to traditional 2D dust maps (D. J. Schlegel et al. 1998), Bayestar19—a 3D dust map—provides more accurate and spatially resolved extinction estimates, effectively mitigating overestimation issues.

Six panels in Figure 6 present the Hertzsprung–Russell (H-R) diagram distributions corresponding to the emission-line profile types, single emission profiles: 1E (without absorption), and 1EA (with absorption), double emission profiles: 2E (without absorption) and 2EA (with absorption), and P Cygni profiles PC and inverse P Cygni profiles IPC, respectively. All diagrams are shown as density plots, where colors transition from purple to red with increasing stellar density.

Stars with the label 1E are broadly distributed across the H-R diagram, with prominent concentrations along both the main sequence and the giant branch. This widespread distribution reflects the diverse physical origins of Hα emission discussed in Section 2, such as stellar envelopes, binary mass transfer, Be star disks, and chromospheric activity in late-type stars (T. Kogure & K.-C. Leung 2007). Essentially, any stellar type capable of producing emission lines may exhibit a single-peaked Hα profile.

In contrast, the 1EA type—also a single-peaked emission profile—shows a distinct concentration in the upper main sequence. This is likely related to emission mechanisms specific to early-type stars. When viewed at certain angles, emission lines from early-type stars may appear embedded in broad and shallow absorption features, producing the characteristic 1EA profile. This interpretation is supported by the catalog of Y.-J. Zhang et al. (2022), in which 1EA is also the most common label among early-type emission-line stars.

Stars with 2E profiles also cluster along the upper main sequence but are shifted further toward the upper left, indicating they are generally hotter, more luminous, and more massive than those with 1EA profiles. A common interpretation is that viewing a Be star disk at an inclined angle results in a narrow absorption core bisecting the emission peak, thus producing a double-peaked profile (T. Rivinius et al. 2013). The observed concentration in the B-type region supports the idea that many 2E stars are classical Be stars. In addition, a distinct concentration is present above the lower main sequence. This concentration lies along the binary sequence, as binary systems are generally brighter than single stars of the same temperature and hence appear slightly elevated relative to the main sequence. At the same time, the spectra of these objects exhibit double-peaked emission features produced by accretion disks, suggesting that this concentration is likely associated with binary systems hosting accretion disks.

The distribution of the rare 2EA profile is more sparse. The limited number of these stars may be due to our stricter selection criteria for this label. Nonetheless, two distinct concentration regions are observed—one at the top of the main sequence and another near the turn-off point toward the red giant branch. These correspond to the typical locations of 2E and 1EA stars, respectively. We propose that the 2EA profile may arise from a combination of double-peaked emission (as seen in 2E profiles) and broad absorption features (as in 1EA), making them detectable in both evolutionary stages.

The distributions of PC and IPC profiles exhibit similar clustering patterns in the H-R diagram. Both of them show varying degrees of concentration along the giant branch and near the main-sequence turn-off. In addition, the IPC profiles display a smaller clustering near the tip of the giant branch. For the distribution of these two types of spectral profiles, we do not yet have a clear understanding and therefore present them here for reference only.

## 8. Summary and Conclusion

Building on the DS, which can efficiently identify line profiles in the large survey, we have identified 45,206 Hα emission-line stars, with 39,497 being newly discovered. In total, 56,649 Hα emission lines have been classified into six categories: single emission profiles (27.1% as 1E and 55.8% as 1EA), double emission profiles (5.3% as 2E and 0.2% as 2EA), and P Cygni-type profiles (7.4% as PC and 4.1% as IPC). Except for certain 1EA type profiles—where severe component blending prevents reliable width estimation—the parameters of each Hα line component, including central wavelength ($\lambda$), width ($w$), and amplitude ($a$), are provided for all stars in the catalog. This constitutes a large catalog that features a substantial number of newly discovered emission-line stars, along with detailed Hα line classifications and associated parameters. It also exhibits the rich cross-identification with existing catalogs, encompassing the broadest range and quantity of potential emission-line stars. Consequently, our catalog includes many candidates worthy of further study and may conceal a significant number of sources with rare or intriguing physical properties.

## Acknowledgments

We gratefully acknowledge financial support from the National Natural Science Foundation of China (grant Nos. 12288102, 12125303, 12090040/3, 12403040, and 11733008), the National Key Research and Development Program of China (grant No. 2021YFA1600401/3), the Yunnan Fundamental Research Projects (grant No. 202101AV070001), the International Centre of Supernovae, Yunnan Key Laboratory (No. 202302AN360001), and the China Manned Space Project (CMS-CSST-2021-A10). We also thank the joint funding from the National Natural Science Foundation of China and the Chinese Academy of Sciences (grant No. U1831125), as well as the Research Program of Frontier Sciences, CAS (grant No. QYZDY-SSW-SLH007). We are particularly grateful for the computational resources provided by the Phoenix Supercomputing Platform at the Yunnan Observatories. Finally, we acknowledge the crucial role of the Guoshoujing Telescope (LAMOST), a National Major Scientific Project initiated and funded by the Chinese Academy of Sciences, and operated by the National Astronomical Observatories, CAS.

## ORCID iDs

Lihuan Yu ● https://orcid.org/0009-0004-7472-0411
Jiangdan Li ● https://orcid.org/0000-0003-3832-8864
Jinliang Wang ● https://orcid.org/0009-0002-1546-8442
Tongyu He ● https://orcid.org/0009-0004-9758-0722
Zhanwen Han ● https://orcid.org/0000-0001-9204-7778